\documentstyle[11pt,paspconf]{article}

\begin{document}

\title{The High Redshift Universe with Adaptive Optics:\\ Recent results from CFHT}
\author{David Crampton}
\affil{Dominion Astrophysical Observatory, National Research Council of Canada,
Victoria, V8X 3X1, Canada.\\ Email: David.Crampton@hia.nrc.ca}

\begin{abstract}
The CFHT Adaptive Optics Bonnette (AOB) has been used to
obtain high spatial resolution (0\farcs1) observations of several
extragalactic targets including the nuclei of nearby galaxies, high
redshift galaxies, AGN, radiogalaxies, the host galaxies of quasars
and gravitational lenses. Examples of these are discussed
and the role of adaptive optics in exploring the high redshift universe
is critically assessed in light of these results.
\end{abstract}


\section{Introduction} The CFHT Adaptive Optics Bonnette (AOB)
provides excellent correction using much fainter guide stars than
previous systems so that observations of many extragalactic targets are
now possible, using either a nearby reference star or the object
itself. A wide variety of extragalactic objects have now been observed
with AOB and some of these are used as examples to discuss some of the
problems encountered and to demonstrate the potential of adaptive
optics for studies of the high redshift universe.

\section{The Adaptive Optics Bonnette} 
The CFHT AOB is based on a curvature wavefront system (Roddier et al. 1991)
with 19 subapertures.  A complete description of the system is given by
Arsenault et al. (1994), and a comprehensive discussion of its
performance by Rigaut et al. (1997).  If the target is to be observed in
the infrared, a dichroic is used to reflect the visible light to
the wavefront sensor (WFS) while transmitting the IR light to the
science detector, but if the target is to be observed in the visible, a
beamsplitter must be used to send a percentage of the light to the
WFS.  The delivered image quality depends on the brightness of the
reference star, its distance from the target, the seeing at the time of
observation and the wavelength of observation. To gain an appreciation
of how these variables affect the images and what can be expected, the
reader is encouraged to visit the ``performance meter'' at
http://www.cfht.hawaii.edu/manuals/aob/psf.html  To achieve FWHM
$\sim$0\farcs12 imaging in the near-IR (diffraction-limited) and  at
$I$ under median seeing conditions, the reference star must be brighter
than R $\sim$14 and within $\sim$30\arcsec\ of the target. However, the
system will provide good correction on stars as faint as R = 17 under
good seeing conditions, and the reference star can be located anywhere
within the 90\arcsec\ diameter field.  AOB offers considerable
advantages in terms of operational efficiency, since it is literally a
push-button operation  that eliminates all the overhead of focussing
and guiding associated with conventional observing.

The observations reported here were all made with the University of
Montreal infrared camera MONICA (Nadeau et al. 1994) which was modified
to give a pixel scale of 0\farcs034 and hence a field of
8\farcs8$\times$8\farcs8.  This very small field is a significant
handicap for many observations, as is the fact that the detector
suffers from persistence problems, i.e., a bright source leaves a
residual image that slowly decays with time.  Since the site seeing is
variable, observations of relatively bright stars usually {\it have} to
be carried out in order to monitor the PSF (point-spread function). The
sequence of observations of very faint sources thus has to be carefully
planned to minimize such problems while adequately monitoring the PSF
to attain the scientific goals.  V\'eran et al. (1997) have recently
demonstrated that statistics of the wavefront sensor signals can be
used to derive an excellent model of the average PSF. This is obviously
ideal in that  the synthetic PSF is simultaneous with the the target
observation and it obviates the requirement of directly observing the
bright star.

\section{Examples of Self-referencing Targets} 
Objects such as the nuclei of nearby galaxies, seyferts and AGN can be
usually guided on directly, i.e., the nucleus is often sufficiently
point-like to enable the WFS to function properly as long as there is
sufficient flux.  For example guiding on the nucleus of M31 was
straightforward, resulting in diffraction-limited images (FWHM =
0\farcs12) of the stars in the nuclear bulge region, despite the fact
that it is double with a fainter component at a distance of
$\sim$0\farcs5. The near-IR image quality is comparable to that of
WFPC on $HST$, enabling accurate magnitudes and colors to be
determined for a significant number of stars in the bulge (Davidge et
al.  1997).  During the commissioning period, other nearby galaxies,
AGN and Seyfert 1 type galaxies with point-like nuclei and magnitudes
in the range m = 11--14 (e.g., bright ones like NGC 4151 to fainter
ones like NGC 6814) were observed with AOB, guiding on the bright
nuclei themselves. However, guiding was  not successful on more diffuse
objects such as the V = 13 nucleus of the Seyfert 2 galaxy Mrk 266, or
the large bright elliptical galaxy which hosts 3C296.

The small MONICA  field  complicates observations of such targets since
neither the sky nor PSF stars are normally included in the frames.  An
accurate estimate of the PSF is essential for these sources in order to
adequately subtract the effects of the bright point source from the
nuclear region, so a nearby reference star of similar brightness must
be monitored throughout the sequence of observations.

\section{Targets with guide stars}
Nearby guide stars must be used for targets which are not sufficiently
concentrated or bright enough for the WFS to perform adequately. In
this case, not only are observations of the guide star itself required
to monitor the PSF variations, but measurements of starfields (e.g.,
globular cluster fields) are also required to estimate the degradation
of the PSF due to anisoplanatic effects.  Steinbring (1997) has developed
semi-empirical software that models this degradation over the field as
a function of the seeing and produces an ``off-axis PSF" that can be
used to help analyse the target data. For some projects this is not
important, but for the observations of quasar host galaxies, for
example, it is vital.  Studies of the latter are notoriously difficult
due both to the faintness of the galaxy itself plus the superposition
of the bright nuclear source. Some quasars have already been
observed with AOB, both in the visible region (at $I$) with a CCD detector and
in the near infrared (mostly at $H$).  Analysis of the z = 1.1 quasar,
1055.3+019, which is well-resolved in both $I$ and $H$, has recently
been published by Hutchings et al. (1998) who suggest that the host
galaxy has been undergoing a close encounter or merger event.

The components of multiply-imaged gravitational lenses formed by
galaxies in the line of sight to distant quasars typically have
subarcsecond separations and consequently their study and monitoring
could significantly benefit from the improved resolution offered by
adaptive optics.  At present, about $\sim$35 such lenses are known
(Kochanek 1997) but virtually none of them have
nearby guide stars that are sufficiently bright to give good
correction. One exception is SBS 1520+530, a
doubly-imaged BAL quasar with a separation of 1\farcs6 which happens to
be only 13\arcsec\ from a m$\sim$12 star. $H$ band images with FWHM =
0\farcs15 taken with AOB reveal the lensing galaxy 0\farcs40 from the
fainter component, offset 0\farcs12 from the line joining components A
and B (Crampton, Schecter and Beuzit 1997).

\section{Registering invisible targets}

Targets which are very faint, particularly those that are diffuse,
often present additional problems since the MONICA field is so small
that there aren't any objects or features that can be used to register
the images. In general, near-IR images have to be dithered to remove the
effects of bad pixels and to improve the flat-field, offsets are often
required to enlarge the area surveyed, and differential flexure between
the WFS and the detector may produce additional shifts. Consequently,
registering and superimposing the images is not trivial. The AOB WFS
coordinates are recorded in the FITS headers and observations of star
fields have been used to calibrate these positions in terms of pixel
location on the detector and/or equatorial coordinates.  Repeated
observations of the same point source (e.g., a quasar) over several
hours show that the WFS positions can be used to register frames with a
dispersion of 1.4 pixels or 0\farcs05. Since the delivered images in
the near-IR usually have FWHM $\sim$0\farcs12, this is clearly
inadequate, so future systems should be designed with a low-order WFS
incorporated close to the detector  to minimize flexure and allow for
sub-pixel ``drizzling" and registration.  Ideally, the coordinates of
this WFS should allow  diffraction-limited images to be registered to
better than $\sim$10\% of their FWHM; for 8m telescopes this will be of
the order of a few mas.

Although the 0\farcs034 pixel scale is necessary to properly sample the
$\sim$0\farcs1 resolution delivered by AOB, many extragalactic projects
would benefit from a larger field. KIR, a new 1K$\times$1K camera with
three times higher sensitivity and a field of 36\arcsec\ has just been
commissioned, so registration will be easier. Even so, astrometry to
the precision that is possible with 0\farcs1 images will be
challenging:  relative positions can be measured to an accuracy of a
few mas, but tying these to an absolute reference frame is difficult.
Since many extragalactic targets are diffuse, a larger pixel scale
would not be a disadvantage for most investigations, and the larger
field would significantly improve the probability of including stars,
providing even better registration and astrometry.

The jet of 3C273 is an example of an object which is barely visible,
even on background-subtracted 300s exposures at $H$ and $K$.  The jet
is located between 11 -- 21\arcsec\ from 3C273 which, of magnitude R =
13, was used as the guide star for AOB.  Given the small field of the
detector, the jet had to be observed at two different locations along
the jet and then all the frames were registered using WFS positions.
Unfortunately the site seeing was poor on average and very variable
(from 0\farcs5 to 1\farcs2) during the observations so, combined
with registration uncertainties, the resulting (preliminary)
image has a resolution of only $\sim$0\farcs3. Observations
with the new KIR camera should be significantly better.

\section{Summary}
Our experience with AOB demonstrates that adaptive optic systems can be
made robust, reliable, efficient and user-friendly and will routinely
deliver 0\farcs1 resolution images from $I$ to $K$ with substantial
gains in the visible. AOB observations of many extragalactic targets
are now feasible and competitive with $HST$.  Images of most of the
objects discussed in this article can be found on the web at
http://www.hia.nrc.ca/science/instrumentation/optical/pueo2/pueo2.html .
Spectroscopic observations using instruments designed to exploit the
AOB image quality are just beginning, although to obtain reasonable
spectral resolution simultaneously with  high spatial resolution will
require adaptive optic systems on 8--10m telescopes for the majority
of extragalactic targets.

\end{document}